**RESEARCH**

# On the importance of laboratory experiments for interpreting exoplanet observations

Maggie A. Thompson[1]

**Abstract**
With the advanced capabilities of ground- and space-based telescopes, exoplanet science is beginning to characterize the physics and chemistry of exoplanet atmospheres. However, interpreting exoplanet observations requires sophisticated modeling tools that rely on laboratory data to provide critical inputs and constraints. In preparation for the influx of observational data that the coming decades will bring, laboratory experiments that simulate the diverse conditions expected in exoplanet atmospheres, surfaces and interiors are vital to advancing models and thereby our understanding of these worlds. Here we discuss the key areas where laboratory experiments are providing fundamental data for exoplanet models, particularly for low-mass planets from rocky worlds to sub-Neptunes. First, we present a series of experiments to measure outgassing and volatile solubilities that are essential for establishing the connection between low-mass exoplanet interiors and their observable atmospheres. We then discuss additional laboratory techniques that can be used to understand the interior-atmosphere connection and simulate the high pressure-high temperature conditions of exoplanet interiors. Next, we summarize the experimental methods used to constrain the spectroscopic properties of atmospheric gases and aerosols along with their formation and reaction mechanisms. We also discuss how similar methods can be used to constrain exoplanet surface compositions, which is important for interpreting observations of atmosphere-less worlds. Finally, we conclude by presenting several examples of astrobiology experiments that constrain how life can modify the atmosphere and surface of rocky exoplanets. Together, these laboratory efforts are crucial to maximizing the scientific yield of exoplanet observations in the coming decades.

## 1 Introduction

Exoplanet science is undergoing an observational revolution, transitioning from the era of exoplanet detection to one of characterization. Over the last 15 years, the space-based Kepler and Transiting Exoplanet Survey Satellite (TESS) missions, along with complementary ground-based observatories, have increased the number of confirmed exoplanets by a factor of ∼20 (Borucki 2018; Winn 2025). The launch of NASA's James Webb Space Telescope (JWST) in 2021 marked the beginning of the era of exoplanet characterization, with JWST providing the most in-depth view of exoplanet atmospheres to date through transmission and thermal emission spectroscopy and direct imaging (Lunine and Bahcall 2025). In the coming years, ground-based Extremely Large Telescopes (ELTs) will achieve first light and characterize exoplanets in reflected light and at high spectral resolution. Exoplanet observations seek to determine these planets' atmospheric compositions and pressure-temperature structures. By combining the measured bulk properties of an exoplanet (e.g., mass, radius and orbital period) with observations of its atmospheric composition and structure, a key goal of exoplanet science is to link the observed atmospheric properties to those of the planet's deeper interior (Lichtenberg et al. 2025). Such efforts are essential to advance our understanding of the physics and chemistry operating in these planets, the most abundant kinds of which we have no analog for in our Solar System (i.e., super-Earths and sub-Neptunes), along with their formation mechanisms.

M.A. Thompson: NHFP Sagan Fellow.

✉ M.A. Thompson
   mthompson@carnegiescience.edu

[1] Earth and Planets Laboratory, Carnegie Institution for Science, Washington, DC 20015, USA

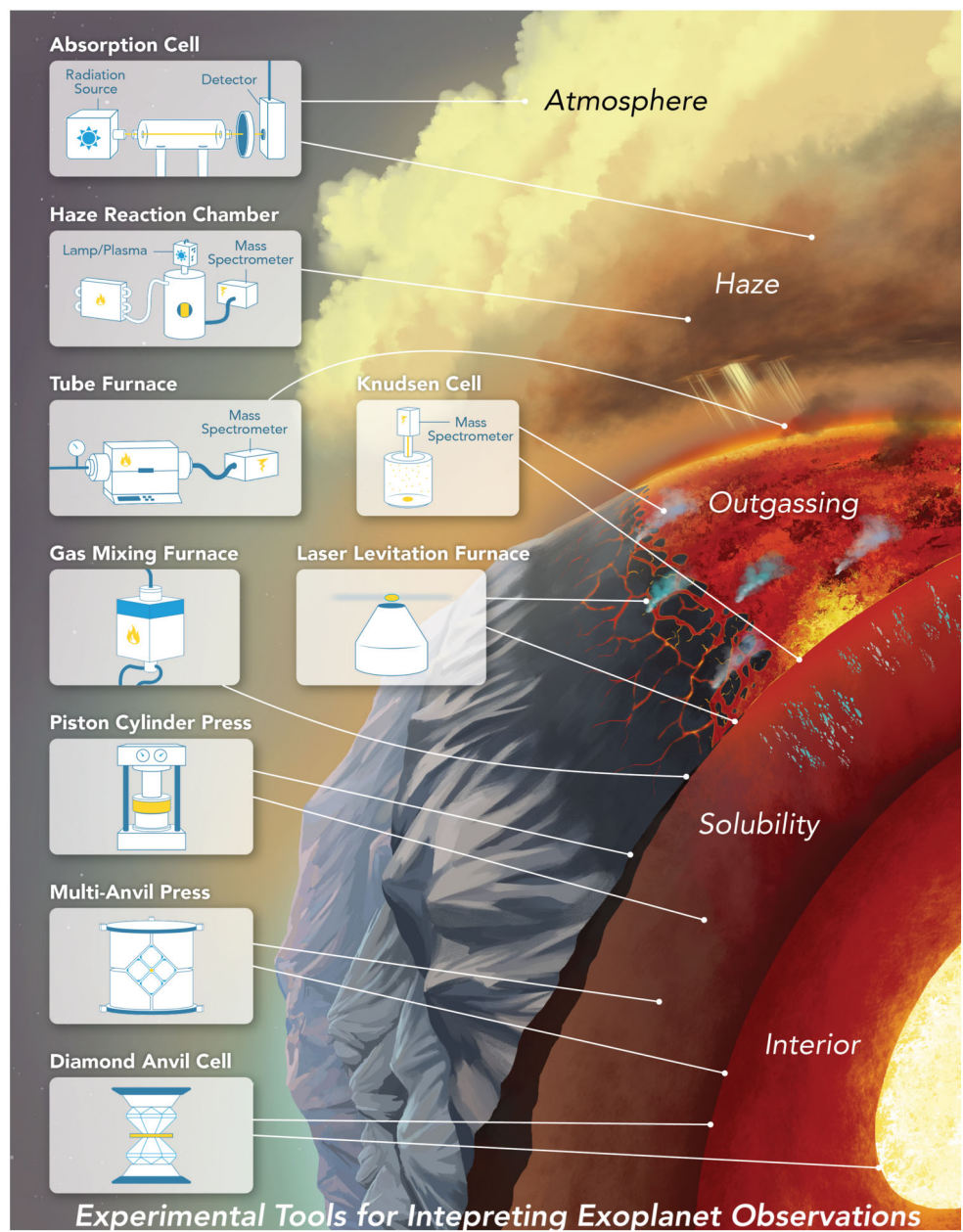

**Fig. 1** Summary of the experimental tools discussed in this review that are used to obtain laboratory measurements that provide fundamental data of atmospheric, surficial and interior properties which are necessary for properly interpreting exoplanet observations. In this review, we categorize these experimental set-ups according to which reservoir of the planet their measurements pertain to: 1) the atmosphere – spectroscopic absorption cells used to measure optical properties of gases and condensates and haze reaction chambers used to create photochemical haze analog materials and study photochemical reactions (Sects. 3 and 4); 2) the surface and interior-atmosphere connection – tube furnaces couped to mass spectrometers, Knudsen cell, gas-mixing furnaces, and laser levitation furnaces (Sect. 2); and 3) the interior – piston cylinder press, multi-anvil press, and diamond anvil cell (Sect. 2). (© 2025 Elena Hartley http://www.elabarts.com)

For the first time, these observations will allow us to place Earth and the planets in our solar system within the wider context of planet formation and evolution in the universe (Lapôtre et al. 2020).

As evidenced by studies of the planetary atmospheres in our solar system, to properly interpret remote atmospheric observations, we need laboratory data to provide critical inputs and constraints for atmospheric models (Fortney et al. 2016). Given the extreme environments in exoplanet atmospheres and interiors expanding to temperatures, pressures, compositions, and radiation environments outside Earth-like conditions, new laboratory experiments that simulate these diverse conditions are needed to provide proper inputs for physical and chemical modeling tools and thereby to correctly interpret the observations (Kohler et al. 2020; Niraula et al. 2022; Potapov and Bouwman 2022; Stapelfeldt and Mamajek 2025; Winiberg et al. 2025). In this review, we summarize the key areas where laboratory data is essential for filling key gaps in exoplanet atmosphere models and for our theoretical understanding of how to link planetary atmospheres to interior properties. This review will focus on laboratory data that is important for exoplanets ranging from rocky, Earth-like worlds to sub-Neptunes, but will discuss spectroscopic and atmospheric chemistry data that also applies to gas giant planets and brown dwarfs. Figure 1 summarizes the main types of laboratory tools discussed in this review that are used to make experimental measurements that provide fundamental constraints on planetary atmospheres, surfaces and interiors which are necessary for properly interpreting exoplanet observations. We present the findings from a series of new experiments that provide critical constraints on how atmosphere-forming volatile elements (e.g., H, C, O, S) partition between the interior and atmosphere of low-mass planets. We also discuss how such experimental data can be best incorporated into atmosphere modeling tools, emphasizing that extrapolations of experimental data to regimes far outside that at which the experiments were calibrated should be done with caution. We highlight specific areas where further laboratory work is needed and anticipated to have a significant impact on existing models.

In Sect. 2, we present laboratory experiments that inform the interior-atmosphere connection of low-mass exoplanets (i.e., rocky worlds and super-Earths) and how these experimental findings are incorporated into modeling tools. The first series of experiments constrain the initial chemical compositions of rocky exoplanet atmospheres that form via outgassing (Thompson et al. 2021, 2023), demonstrating that remnant planet-forming materials in our solar system, as represented by meteorite samples, can inform the phase space of initial atmospheric compositions of rocky exoplanets. The second type of laboratory experiments focuses on determining the solubility of volatile elements into the hot, molten interiors of low-mass exoplanets, which is essential for understanding the atmospheres of highly irradiated exoplanets termed 'magma worlds.' We will present the findings from new water solubility experiments which we used to derive a melt composition- and temperature-dependent $H_2O$ solubility law which provides important insights into nebular ingassing on young rocky planets and magma worlds as a storage mechanism for hydrogen (Thompson et al. 2025). We then discuss two additional high-temperature experimental set-ups that are essential for connecting low-mass planet interiors and atmospheres. We conclude this section by briefly summarizing high pressure-temperature experiments that are needed to properly derive mass-radius relationships for exoplanets of various bulk compositions and to model the interior structure of exoplanets, especially that of super-Earths and sub-Neptunes.

In Sect. 3, we summarize the laboratory tools used to obtain spectroscopic and atmospheric chemistry data which provide critical inputs on gas opacities, cross sections, and thermo- and photo-chemical reactions. In Sect. 4, we present laboratory work to simulate the formation and chemical and physical properties of atmospheric aerosols, both photochemical hazes and cloud condensates, and highlight the need to simulate their formation under diverse radiation environments. Section 5 discusses spectroscopic measurements of exoplanet surface analog materials and their importance for interpreting observations of airless exoplanets.

In Sect. 6, we briefly discuss astrobiology laboratory experiments that are important for assessing potential exoplanet biosignatures with next-generation missions like the Habitable Worlds Observatory and the Large Interferometer for Exoplanets. Finally, we conclude in Sect. 7.

## 2 Connecting the interiors and atmospheres of low-mass exoplanets

For low-mass ($\lesssim 15$ $M_{Earth}$) exoplanets, their atmospheric composition and structure are strongly influenced by interior and surficial processes. Such planets can form their atmospheres through two main mechanisms: 1) primary atmospheres, which form via accreting $H_2$- and He-dominated gas from the stellar nebula and 2) secondary atmospheres, which form via outgassing during and after planetary accretion (Elkins-Tanton and Seager 2008; Lammer et al. 2018). For a low-mass planet to retain a primary atmosphere early in its evolution, it must accrete enough mass within the lifetime of the stellar nebula gas disk. During a rocky planet's formation, heat is generated in its interior primarily due to accretion (i.e., gravitational energy), core formation and, depending on its astrophysical formation environment, possibly radioactive decay of short-lived major element isotopes (e.g., Gaillard et al. 2021; Lichtenberg et al. 2022). These thermal processes in the interior shape the outgassing composition of its earliest secondary atmosphere and likely generate global magma oceans on these planets' surfaces (Chao et al. 2021). As a planet's interior cools over time, its secondary atmosphere continues to be controlled by interior processes, including magmatic outgassing, melt- or solid-state convection and geochemical cycling. It is also important to note that some of the low-mass exoplanets discovered to date are highly-irradiated on ultra-short periods ($\sim$hours to days) and can likely sustain magma oceans on their surfaces for prolonged periods of their history. Experimental constraints on both the outgassing compositions expected for young, rocky planets and how major atmosphere-forming gases can dissolve into magma oceans is critical to our understanding of how the interiors and atmospheres of low-mass exoplanets are connected.

### 2.1 Outgassing experiments

Extensive theoretical work has been done to model planetary outgassing, primarily for Earth and the other rocky planets in the solar system, but also for low-mass exoplanets (e.g., Abe and Matsui 1985; Zahnle et al. 1988; Schaefer and Fegley 2007, 2010; Gaillard and Scaillet 2014; Lammer et al. 2018; Herbort et al. 2020). However, experimental data to constrain these models has been limited. One way to inform these models is to directly or remotely measure the outgassing compositions from various geothermal sites on Earth (e.g., volcanoes, hydrothermal vents) or on other Solar System rocky bodies (e.g., volcanoes on Io, cryovolcanic plumes on Enceladus and Europa). Similarly, analytical methods can determine the volatile content of planetary materials that were subject to varying amounts of thermal processing to infer outgassing conditions (e.g., Newcombe et al. 2023). In the lab, there are various instrumental techniques that can be used to study evaporation and outgassing processes, including mass spectrometry, thermogravimetric analysis, gas chromatography, infrared spectroscopy, shock-induced devolatilization and vapor deposition (e.g., Gooding and Muenow 1976; Court and Sephton 2009; Braukmüller et al. 2018; Springmann et al. 2019). However, many of the prior laboratory outgassing experiments were limited by the number of samples studied, the temperatures to which the samples were heated, and the number of gas species accurately measured. Therefore, we have started to fill this gap by performing a series of outgassing experiments on a set of exoplanet-analog materials.

While the range of plausible bulk compositions for low-mass exoplanets is vast, we can use the solar system's rocky bodies as a useful guide. Earth, Venus and Mars are believed to have formed via accretion of planetesimals with bulk compositions similar to those of chondritic meteorites, which were also a likely source of their atmospheric volatile elements (e.g., Lodders 2000; Lammer et al. 2018). Chondrites come from undifferentiated planetesimals, indicating that their parent bodies did not experience sufficient heating to have melted, and therefore, they serve as representative samples of the primitive planet-forming material in the solar nebula and likely that around other Sun-like stars. It is important to note that recent works have shed some doubt on the idea that the solar system's rocky planets can have their bulk compositions explained by meteorite samples alone (e.g., Burkhardt et al. 2021; Sossi et al. 2022). Nevertheless, since chondritic meteorites are some of the only samples that preserve the composition of aggregate material during planet formation around Sun-like stars and are available for laboratory study, they are essential samples to study experimentally to inform the connection between rocky planet interiors and their atmospheres.

To fill the present gap in experimental constraints on the outgassing compositions from chondritic meteorites, which serve as potential exoplanet-analog materials, we performed a series of outgassing experiments on carbonaceous (CM-type) chondrites. In the first study, we heated three CM chondrites in a vacuum ($\sim 10^{-9}$-$10^{-8}$ bar) tube-furnace connected to a residual gas analyzer (RGA) mass spectrometer from 475 to 1475 K (Thompson et al. 2021) (Fig. 2). As each sample was heated, the partial pressures of the outgassing molecular species were continuously monitored with the RGA mass spectrometer. As Fig. 2 demonstrates, these experiments determined that $H_2O$, $H_2$, CO, $CO_2$, and $H_2S$ outgas from CM chondrites, with water being the most abundant outgassing species over the entire temperature-range ($\sim 66\%$), followed by CO ($\sim 18\%$) and $CO_2$ ($\sim 15\%$) and smaller amounts of $H_2$ and $H_2S$ (up to $\sim 1\%$) (where the percentages are the abundances given by the partial pressures normalized to the total pressure of all measured released gases summed over temperature and expressed as percentages). These experiments simulate open-system outgassing conditions whereby the composition of the sample changes as it is heated and volatiles are degassed and removed.

By comparing these experimental results to the predicted outgassing compositions using a thermochemical equilibrium model, we find that while $H_2O$, CO and $CO_2$ constitute substantial fractions of the vapor phase over the explored temperature range in both cases, there are also important differences. For example, $H_2S$ outgasses at higher temperatures ($\sim 1200$ K) in our experiments compared to that predicted by the equilibrium model ($\sim 900$ K), which could be due to chemical kinetics effects that inhibit the necessary phase change for sulfur to outgas. In addition, the amount of $H_2$ gas that outgasses during the experiments is much less than that predicted by chemical equilibrium models, which is likely due to disequilibrium of gas phase reactions involving hydrogen in the experiments. This study provides important experimental constraints on a low-mass planet's initial outgassed atmospheric composition assuming its bulk composition is CM chondrite-like, revealing that $H_2O$-rich steam atmospheres with significant amounts of CO and $CO_2$ are likely to form under these low-pressure conditions and at temperatures up to $\sim 1500$ K (Thompson et al. 2021).

In a second study, we determined the outgassing trends for a suite of moderately volatile and major elements in CM chondrites (e.g., Fe, Mg, Zn, S, P) and explored the effect of atmospheric pressure and redox state by performing heating experiments under vacuum conditions ($\sim 10^{-8}$ bar) and at 1 bar in air. To quantify outgassing of these other elements, we analyzed the samples after the heating experiments using inductively coupled plasma mass spectrometry (ICP-MS). Of the heavier elements measured, we find that both sulfur and zinc significantly outgas at temperatures above $\sim 1000$ K under vacuum whereas only sulfur outgasses between $\sim 1000$-$1300$ K under atmospheric pressure and more oxidizing conditions (Thompson et al. 2023). Through these experiments, we demonstrated that outgassing trends from planetary analog materials depend on the mineral host phases and how the breakdown of these phases is affected by the surface pressure and redox environment. As in Thompson et al. 2021, this second study also finds important differences in the outgassing trends measured experimentally and those predicted by chemical equilibrium models, which are primarily due to the open-system nature of the outgassing

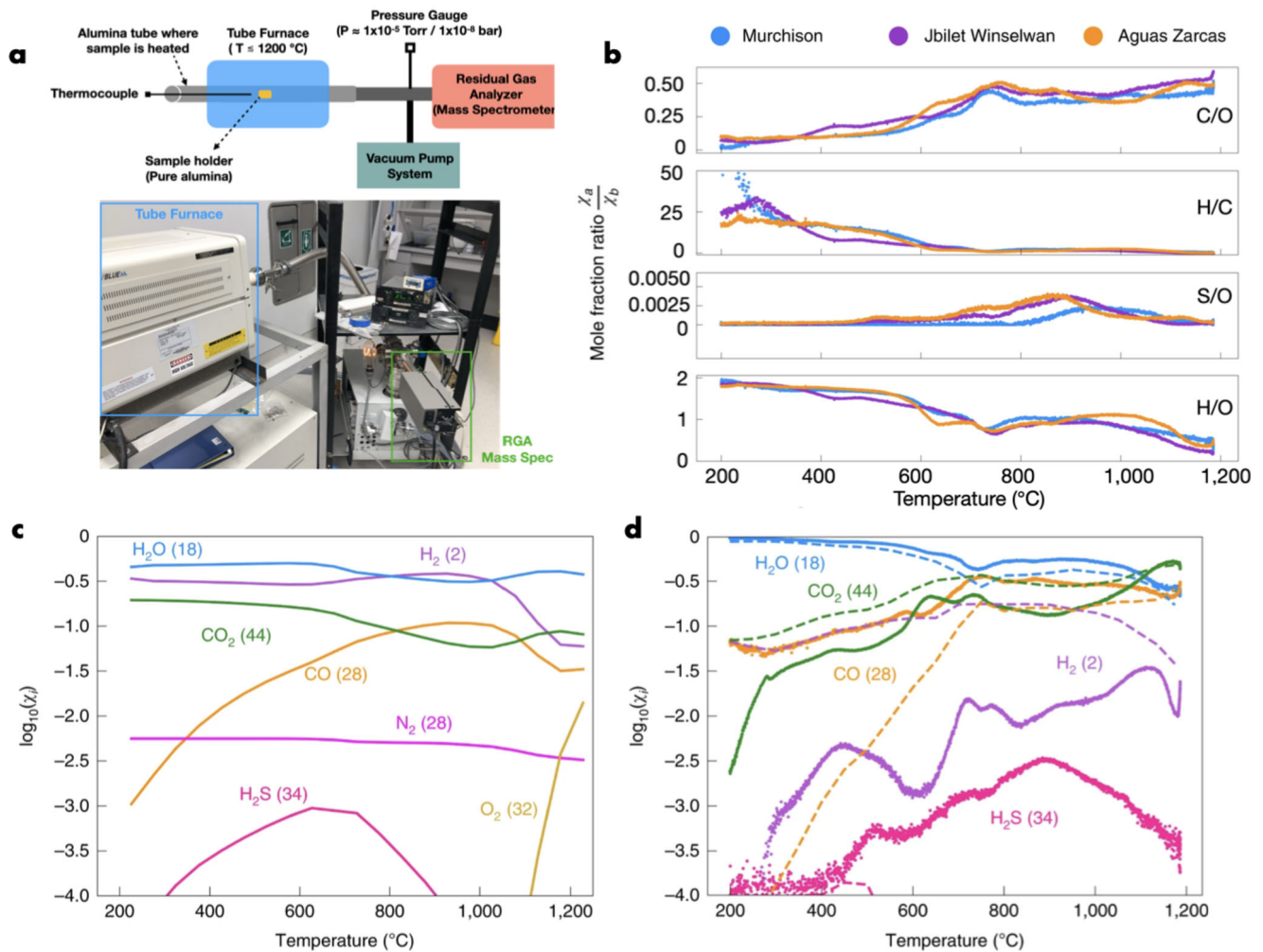

**Fig. 2** a) Experimental set-up to perform the outgassing experiments of Thompson et al. 2021, 2023, (b) the experimentally-measured ratios of outgassed volatile elements (H, C, O, S) as a function of temperature for the three CM chondritic meteorite samples measured in Thompson et al. 2021, (c) the average outgassing composition from the three CM chondrites predicted by chemical equilibrium calculations and (d) measured experimentally as a function of temperature (b, c and d from Thompson et al. 2021) experiments compared to the closed-system models along with chemical kinetics effects that take place in the experiments but are not included in the models (Thompson et al. 2021, 2023). Together these two studies demonstrate that if a planet's bulk composition is similar to that of CM chondrites, then at temperatures up to ∼1300 K and assuming a low surface pressure (< 1 bar), its initial outgassed atmospheric composition will be composed of H, C, O, and S (and possibly Zn if the surface pressure is ≪ 1 bar), providing experimentally-informed surface boundary conditions for models of low-mass exoplanets' early atmospheres.

These outgassing experiments are already being used to inform models of the atmospheric compositions of known rocky planets within and beyond the solar system. For example, atmospheric retrieval codes like HyDRo used these experiments to inform which molecular species to include in the model for interpreting temperate ($T_{eq}$ ∼ 400-800 K) rocky exoplanet observations (Piette et al. 2022). Similarly, in preparation for JWST observations, atmospheric models of the rocky exoplanet GI 486 b used the experimental outgassing results to inform which gas species could be the dominant constituents in its atmosphere (Caballero et al. 2022). These experiments are also useful for constraining the composition of Venus's early atmosphere that formed via outgassing (Salvador et al. 2023). Forthcoming experiments will measure the outgassing compositions from a wider set of chondritic meteorite samples, including ordinary and enstatite chondrites, which will reveal how outgassing compositions vary with planetary bulk composition and redox state. Vapor deposition experiments on chondritic samples using heated evacuated silica-glass tubes will also provide important constraints on the types of cloud condensates that form in outgassed atmospheres (Anzures et al. 2026), which will provide some of the first experiments to compare with models of anticipated exoplanet condensates (e.g., Mbarek and Kempton 2016).

In addition to meteorite samples, we can also generate synthetic exoplanet analog materials based on the measured chemical compositions of exoplanet host stars, since the stellar composition is believed to be a first-order proxy for the bulk silicate composition of its exoplanet (e.g., Hinkel et al. 2014; Hinkel and Unterborn 2018; Brugman et al. 2021; Guimond et al. 2024). For example, the Hypatia catalog provides stellar abundance data for stars in the solar neighborhood (Hinkel et al. 2014, https://www.hypatiacatalog.com/). We can also use the observed chemical compositions of "polluted" white dwarfs, which are remnant stellar cores once a star sheds its outer layers after the red giant phase that have elements heavier than He in their spectra. These heavier elements are from accretion of formerly orbiting rocky bodies, and therefore these measurements provide important insights into the chemical composition of rocky exoplanetary material (e.g., Doyle et al. 2019; Putirka and Xu 2021). Future experimental efforts should utilize constraints from the compositions of both exoplanet host stars and polluted white dwarfs to synthesize plausible exoplanet compositions and then analyze their outgassing compositions over a range of experimental conditions.

## 2.2 Solubility experiments

Most rocky planets likely undergo at least one magma ocean phase during their early evolution, and, of the low-mass, rocky exoplanets discovered to date, a subset of this population exists in highly irradiated environments close to their host stars such that they can sustain magma oceans on their surfaces (Chao et al. 2021). These "magma worlds" are particularly amenable to near-term characterization with *JWST* due to their observational advantage of having short orbital periods and bright dayside infrared fluxes. Many, if not most, magma exoplanets are tidally-locked (i.e., orbital period matches rotation period), having permanent day- and night-sides which means that the dayside likely possesses a magma ocean but the nightside conditions depend on the presence of an atmosphere that could redistribute heat across the planet. Sub-Neptune exoplanets may also harbor magma oceans underneath their thick envelopes, and such a scenario has already been found to be consistent with recent *JWST* observations of two sub-Neptunes, K2-18 b and TOI-270 d (Shorttle et al. 2024; Nixon et al. 2025).

Given the likely prevalence of magma oceans among the known exoplanet population, it is necessary to understand how they impact the observable atmosphere. The atmospheric composition overlying a magma ocean depends primarily on 1) the planet's bulk inventory of atmosphere-forming elements (e.g., H, C, N, O, S) (as discussed in Sect. 2.1) and 2) the solubilities of these gases in the magma ocean, which in turn depend on multiple factors such as the planet's surface pressure and temperature at the magma ocean-atmosphere interface and the melt composition. Solubilities of volatile species (e.g., $H_2$, He, $H_2O$, $CO_2$, CO, $CH_4$, $SO_2$, $N_2$) are largely determined via laboratory experiments, and most prior efforts have focused on understanding their solubilities in Earth-like silicate melts (e.g., basalt, andesite, rhyolite) under moderate to high-pressure conditions relevant to Earth's crust and mantle (∼100s of bars to a few GPa) (e.g., Dixon et al. 1995; Mysen et al. 2009; Hirschmann et al. 2012; Ardia et al. 2013; Yoshioka et al. 2019; Dasgupta et al. 2022; Foustoukos 2025). A variety of experimental techniques are used to conduct solubility experiments including gas-mixing furnaces (P: ∼1 bar, T: up to ∼1900 K), laser-levitation heating (P: ∼1 bar, T: up to ∼5000 K), internally- or externally-heated pressure vessels (P: up to ∼0.6 GPa, T: up to ∼1500 K), piston cylinder presses (P: ∼0.5-4 GPa, T: up to ∼2000 K), multi-anvil presses (P: ∼4-27 GPa, T: up to ∼2800 K), and diamond anvil cells (P: up to ∼300 GPa, T: up to ∼6000 K) (Fig. 1). Most of these methods require ex-situ analysis to quantify the concentration of a dissolved volatile element and to determine its speciation in the melt. Therefore, a critical aspect of these experiments is to preserve the melt composition by quenching it to a glass. The effects of quenching on the speciation (and possible diffusion) of the dissolved component has been and continues to be an active area of study and subject of debate (e.g., Solomatova et al. 2020). Common analytical techniques used for quantifying solubility include Fourier-Transform infrared spectroscopy (FTIR), Raman spectroscopy, secondary ion mass spectroscopy (SIMS), nuclear magnetic resonance (NMR) spectroscopy, electron probe micro-analyzer (EPMA) and X-ray absorption near-edge structure spectroscopy (XANES).

Under chemical equilibrium conditions, which are expected if a planet's mantle mixing timescale is fast relative to its atmospheric cooling timescale, the concentration of a dissolved volatile species will be proportional to the fugacity of the conjugate vapor species in the gas phase (following Henry's law) (or its fugacity raised to some power). Solubility laws are often determined by fits to experimental data and informed by the thermodynamics of the governing dissolution reaction (e.g., Newcombe et al. 2017; Thomas and Wood 2021; Boulliung and Wood 2022; Dasgupta et al. 2022; Thompson et al. 2025). However, there is currently limited experimental data on the solubilities of major atmosphere-forming gases in diverse melt compositions unlike those found in modern Earth's mantle, including primitive compositions representative of Earth's magma ocean stage (e.g., peridotite) and that of the other solar system rocky planets, along with exoplanet-analog compositions. In addition, solubility measurements conducted at low total pressures (∼1-200 bar) and extremely high pressures

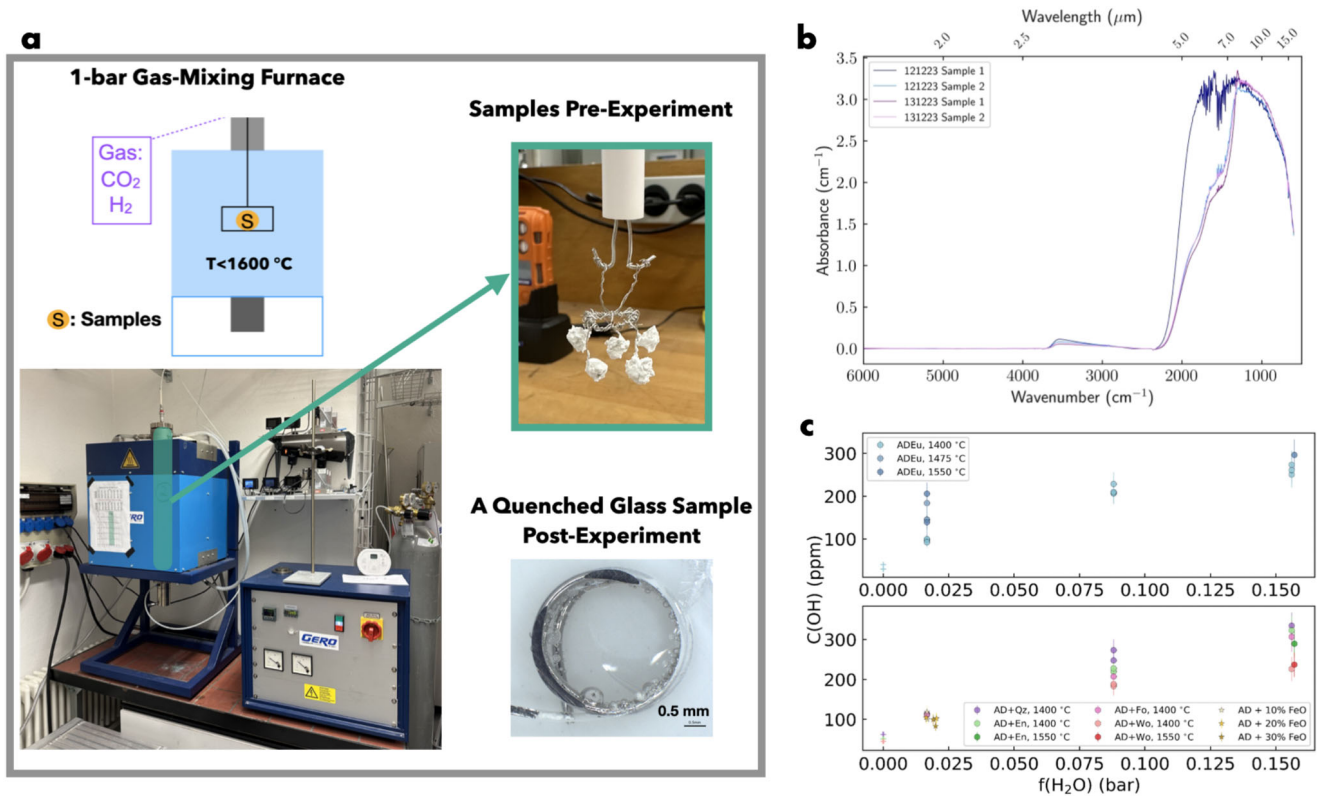

**Fig. 3** (a) Experimental set-up of the gas-mixing furnace used to perform the water solubility experiments of Thompson et al. 2025, including photos of the silicate samples before and after an experiment (modified from Thompson et al. 2025), (b) FTIR transmission spectra for four of the silicate glasses, showing the 3550 cm$^{-1}$ absorption feature due to OH from Thompson et al. 2025, and (c) dissolved concentrations of OH in silicate glasses (ppm) as a function of the H$_2$O fugacity (bar) from Thompson et al. 2025 and temperatures (>30 GPa and >3000 K) relevant for the magma ocean-atmosphere boundary of rocky exoplanets and that of more massive sub-Neptunes, respectively, are presently lacking.

To fill the gap in low-pressure solubility data for H$_2$O, we conducted a series of experiments using a 1-bar H$_2$-CO$_2$ gas-mixing furnace on various silicate melts under reducing atmospheric conditions ($\Delta$IW between $-4$ and $-1$) and temperatures from 1673 to 1823 K (Thompson et al. 2025). To quantify the atmospheric oxidation state, oxygen fugacity (f(O$_2$)), which represents the chemical potential of oxygen in a system, is expressed in log10 units relative to the iron-wüstite buffer (i.e., $\Delta$IW) from Hirschmann (Hirschmann 2021). Constraints on water and other H-bearing species' solubilities are particularly important for considering the origin of volatiles for low-mass planets, as ingassing of a primary (i.e., nebular) H$_2$-rich atmosphere is a proposed mechanism for bringing hydrogen to rocky planets, including early Earth (e.g., Olson and Sharp 2018, 2019; Young et al. 2023). Our experiments measured water solubility in a range of silicate melt compositions in the Ca-Mg-Al-Si-Fe-O system and spanned a range of H$_2$ and H$_2$O fugacities: f(H$_2$) from 0.68 to 0.98 bar and f(H$_2$O) from 0.02 to 0.16 bar. We quantified the amounts of dissolved H$_2$O by transmission FTIR spectroscopy, using the intensity of the 3550 cm$^{-1}$ absorption feature due to the OH$^-$ stretching band (Fig. 3) and the Beer-Lambert law:

$$X(OH) = \frac{I_{3550} M(OH)}{d \rho \varepsilon_{3550}}$$

which relates the mole fraction of dissolved OH in the glass (X(OH)) to the absorbance of the 3550 cm$^{-1}$ band produced by the OH stretching mode (I$_{3550}$), its molar mass (M(OH)), the path length of light transmitted through the absorbing medium (d), the glass density ($\rho$), and the molar absorption cross section for the 3550 cm$^{-1}$ band ($\varepsilon_{3550}$).

We found that under these low-pressure conditions water dissolves in silicate melts exclusively as OH$^-$, with concentrations ranging from $\sim$83 to 330 ppm OH, and that the concentration of dissolved OH increases with f(H$_2$O)$^{0.5}$ (Thompson et al. 2025). Despite conducting experiments under reducing conditions with relatively large H$_2$ fugacities, we did not detect any dissolved molecular H$_2$ in our glasses, which is expected if we assume that H$_2$ dissolves via the same physical mechanism as noble gases (Carroll and Stolper 1993). By extrapolating the observed relation-ship between atomic/molecular diameter and solubility for noble gases to $H_2$, the predicted dissolved $H_2$ concentrations should be $<10^{-2}$ ppm, multiple orders of magnitude less than the amount of dissolved OH. However, at higher pressures, $H_2$ dissolution becomes more important (Hirschmann et al. 2012; Foustoukos 2025). We determined that the governing dissolution reaction for water in silicate melts at these low-pressure conditions is:

$$H_2O(g) + O^{2-}(l) = 2OH^-(l)$$

where its equilibrium constant ($K_{eq}$) is given by: $K_{eq} = \frac{a(OH^-)^2}{f(H_2O)a(O^{2-})}$. Assuming an ideal solution, the activity of OH is equivalent to the mole fraction: $a(OH^-) = X(OH^-)$, and we can therefore relate the dissolved concentration of OH to $f(H_2O)^{0.5}$.

We used Bayesian parameter estimation to derive a new water solubility law based on both our new experimental data and that of two prior studies of water solubility at 1-bar that explored different temperatures and melt compositions, including lunar basalt and peridotite (Newcombe et al. 2017; Sossi et al. 2023). Our new water solubility law includes the effects of melt composition, temperature and $f(H_2O)$ and is given by:

$$\begin{aligned} X(OH) &= 7.19E-4 \\ &\times \exp(\frac{-1511.1 X_{CaO} + 886.5 X_{SiO_2} - 1015.2 X_{MgO} + 890.6 X_{Al_2O_3} - 1755.9 X_{FeO}}{T}) \\ &\times f(H_2O)^{0.5} \end{aligned}$$

Where $X_i$ are the mole fractions of the various oxide components in the melt ($CaO$, $SiO_2$, $MgO$, etc.), and the standard deviations of the parameters' posterior distributions are given in Table 4 of Thompson et al. (Thompson et al. 2025). This water solubility law fits the experimental data of Newcombe et al. 2017; Sossi et al. 2023 and Thompson et al. 2025 well with an $r^2$ of 0.93. We recommend that this law be used to assess water solubility at pressure conditions below ∼1 kbar and temperatures between 1600-2100 K, relevant for the magma ocean surface conditions of low-mass, rocky planets. In general, it is essential to note the conditions at which solubility experiments are conducted to ensure that their corresponding solubility laws are not being applied to extrapolated regions far beyond their calibrated pressure-temperature-composition space.

We applied this model to simulate the amount of hydrogen that could ingas into a rocky planet's magma ocean as a function of its surface pressure at the magma ocean-atmosphere interface and its atmospheric oxygen fugacity. We determined that a 1 $M_{Earth}$ planet with ∼300 bar surface pressure, set by accretion of solar-like nebular gas and a relatively oxidized overlying atmosphere of $\Delta IW \sim -3$, can ingas ∼100 ppm of hydrogen (see Thompson et al. 2025, Fig. 6). While this amount of ingassed hydrogen matches current estimates of the hydrogen content in the bulk silicate Earth, there are multiple issues with Earth's hydrogen being sourced entirely from nebular ingassing. For example, Earth's core formation age post-dates the dissipation of the nebular gas disk (Kleine and Walker 2017), Earth's surface has near-chondritic D/H ratios (Alexander et al. 2012), and outgassing and subsequent escape of hydrogen will result in a lower amount of hydrogen retained in Earth's interior over time. Nevertheless, nebular ingassing can be an important source of volatiles for low-mass exoplanets, as population studies suggest that most rocky exoplanets captured and retained nebular atmospheres for significant periods of time (Ginzburg et al. 2018). Our work highlights the influence of oxygen fugacity at a planet's magma ocean surface in controlling the amount of hydrogen that can dissolve into the interior. For example, a 1 $M_{Earth}$ planet with a magma ocean surface of $\Delta IW = 0$ can dissolve as much hydrogen as a 4 $M_{Earth}$ planet at more reducing conditions of $\Delta IW = -4$ (Thompson et al. 2025).

### 2.3 Incorporating experimental constraints into coupled interior-atmosphere models

Models seeking to accurately predict the atmospheric compositions of low-mass exoplanets overlying magma ocean surfaces must include the effects of volatile dissolution into the molten interior. To this end, we recently developed an open-source volatile partitioning Python package called *Atmodeller*, which computes the chemical equilibrium conditions at a planet's surface-atmosphere interface including condensed (both melt and solid) phase thermodynamics, compositional-dependent volatile solubilities into melts, gas non-ideality at high pressures, and condensation of $H_2O$ (liquid), carbon (as graphite) and sulfur (as $\alpha$-S) (Bower et al. 2025). To demonstrate the effects of volatile solubilities on the atmospheric composition at a planet's magma ocean-atmosphere interface, we ran a set of simulations with *Atmodeller* on a young TRAPPIST-1 e analog to serve as an example of a rocky exoplanet in its magma ocean phase. We assumed a young TRAPPIST-1 e had a surface temperature of 1800 K and fully molten mantle, and we ran two sets of Monte Carlo simulations each with 10,000 realizations, where we uniformly sampled (in $\log_{10}$ space) the total hydrogen and carbon inventories of the planet and along with its oxygen fugacity (between IW-5 and IW+5) (see (Bower et al. 2025) for details on the simulations). In the first simulation, we ignore the effects of volatile solubilities into the planet's interior, whereas in the second simulation we include $H_2O$, $H_2$, CO, $CO_2$, $CH_4$, $N_2$, S-bearing species solubilities, assuming a young TRAPPIST-1 e had a basaltic magma ocean composition.

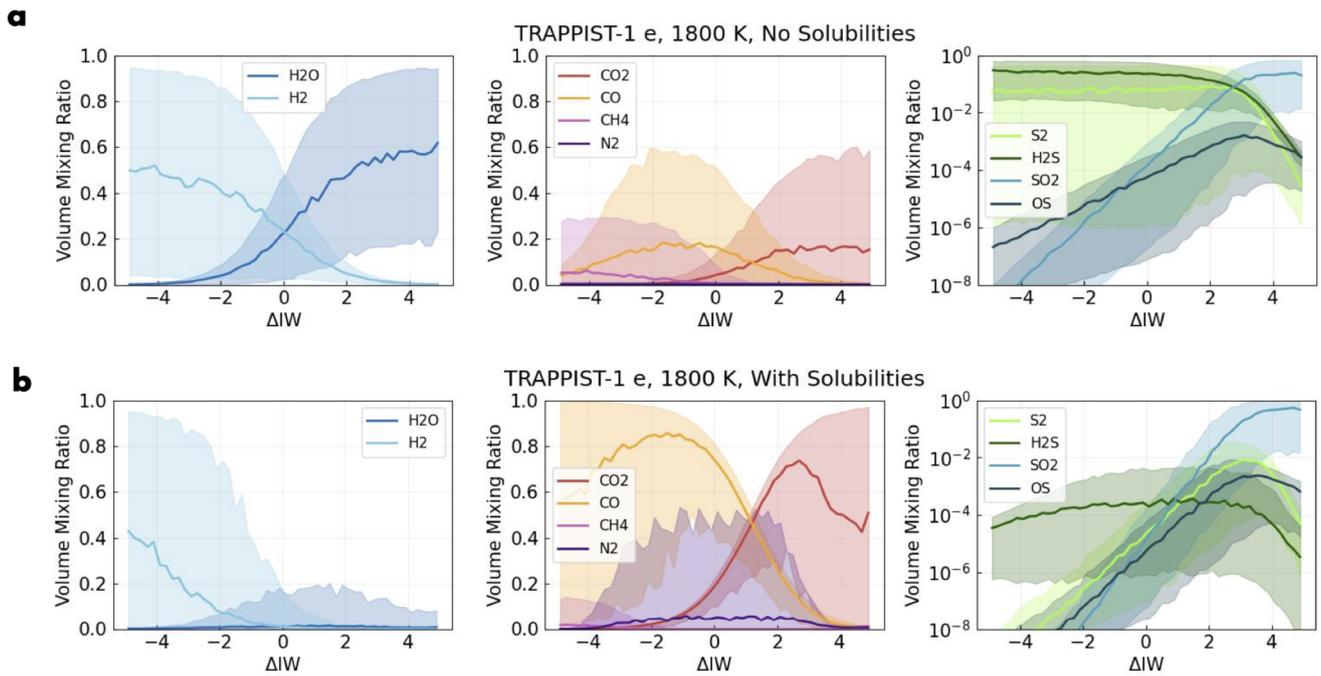

**Fig. 4** The effects of including volatile solubilities on the atmospheric composition at the magma ocean surface ($T_{surf} = 1800$ K) of a young TRAPPIST-1 e analog as an example of a rocky exoplanet in the magma ocean phase. (a) The volume mixing ratios (VMR) of $H_2$ and $H_2O$ (left), $CO_2$, $CO$, $CH_4$ and $N_2$ (middle) and sulfur species (right) at the surface as a function of oxygen fugacity ($\Delta$IW) in the case where solubility of these volatiles into the interior is not included. (b) The same as above except volatile solubilities are included assuming a basaltic melt composition. The solid lines show the mean VMR for a given gas species binned over the explored $f(O_2)$ range of the simulation, and the light shaded region is the $1\sigma$ standard deviation of the simulated data. As described in the text, models are generated with *Atmodeller* and are from Bower et al. 2025

As Fig. 4 illustrates, the atmospheric composition at the surface of a young TRAPPIST-1 e is strongly influenced by dissolution of major atmosphere-forming gases into its interior. If volatile solubilities are ignored, TRAPPIST-1 e's atmospheric composition at the surface will be dominated by H-bearing species, $H_2$ under reducing conditions ($<$IW-2) or $H_2O$ under more oxidizing ones ($>$IW+2) with smaller proportions of carbon species that vary with the planet's redox state (Fig. 4a). However, if volatile solubilities are included, the surface atmospheric composition will instead be dominated by carbon species, CO under reducing conditions ($<$IW-1, with $H_2$ also being abundant at the most reduced conditions below $\sim$IW-4) and $CO_2$ under oxidizing conditions ($>$IW+2, with $SO_2$ becoming a major atmospheric component at the most oxidizing conditions above $\sim$IW+4) (Fig. 4b). This difference is largely due to the high solubility of water in silicate melts, which reduces the abundances of H-bearing species in the atmosphere (Bower et al. 2025). Therefore, volatile solubilities have a strong influence on the atmospheric compositions of low-mass exoplanets with molten surfaces. To properly interpret current and upcoming observations of magma exoplanets (e.g., 55 Cancri e, K2-141 b, TOI 561 b), atmosphere models must include the impact of volatile solubilities in setting the surface atmospheric compositions of these worlds.

We recommend that *Atmodeller*, which includes a library of experimentally-calibrated solubility laws for various atmosphere-forming molecules (Fig. 5 and see Table 1 in Bower et al. 2025), and other coupled interior-atmosphere volatile partitioning models should be used to aid in interpreting magma exoplanet observations. For example, recent work by Seidler et al. 2025 utilizes a coupled interior-atmosphere modeling framework (including *Atmodeller*) to account for both the equilibrium vaporization of mineral gases and magma ocean-atmosphere volatile partitioning to assess how observations of magma exoplanet atmospheres (e.g., 55 Cancri e, Hu et al. 2024) can be linked to their interior chemical compositions (Seidler et al. 2025). As noted above with our new $H_2O$ solubility experiments, it is important that models consider the experimental calibration range over which volatile solubility laws are derived (Fig. 5). To best prepare for upcoming magma exoplanet observations, future experimental efforts should focus on quantifying the effects of melt composition (including diverse compositions unlike modern Earth) and high pressures and temperatures on the solubilities of major atmosphere-forming gases.

Low-mass exoplanet atmospheric observations probe the upper levels of these planets' atmospheres on the order of mbar pressures (Greene et al. 2016). Therefore, models seeking to interpret atmospheric observations of rocky ex-oplanets must account not only for volatile solubilities, but also how the atmospheric composition changes from the surface to the higher, observable regions due to changes in temperature and pressure along with thermochemical and photochemical processes (e.g., Liggins et al. 2023; Watanabe and Ozaki 2024, and as discussed further in Sect. 3). Coupling interior-atmosphere models like *Atmodeller* to radiative transfer, thermo- and photochemical models is essential for linking atmospheric observations to properties of these planets' surfaces and interiors, such as planetary redox state and bulk composition.

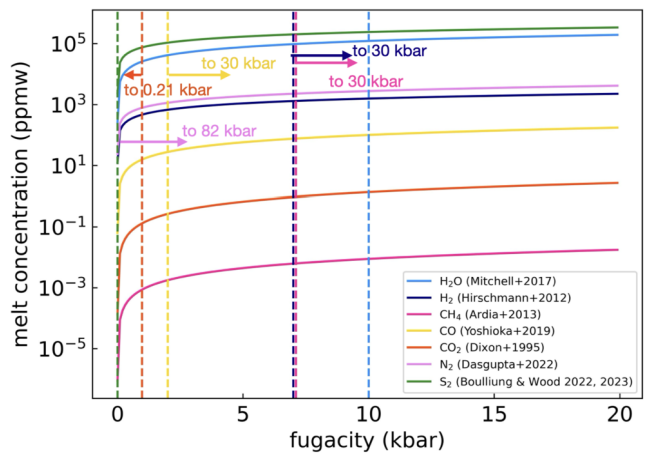

**Fig. 5** Example of some of the volatile solubility laws included in *Atmodeller* (Bower et al. 2025), showing the dissolved concentrations of various volatile species in basaltic melt (ppmw) as a function of that species's fugacity in the gas phase (kbar). These relationships assume a constant surface temperature of 1700 K and a total atmospheric pressure of 50 kbar. If the solubility law also depends on oxygen fugacity, we have set it equal to that of the iron-wüstite buffer at 1700 K and 50 kbar from Hirschmann 2021. The vertical dashed lines indicate the experimental pressures at which the solubility laws were calibrated. For those studies that were calibrated on experiments that spanned a range of pressures, the arrows indicate the range

### 2.4 High-temperature experimental tools for connecting exoplanet interiors and atmospheres

In addition to the experimental methods discussed in Sects. 2.1 and 2.2 to conduct outgassing and solubility experiments, two novel laboratory instruments are currently providing (or under development to provide) fundamental constraints on the interior-atmosphere connection for low-mass exoplanets at high temperatures. The first system is an aerodynamic laser-heated levitation furnace (ALLF), which is used to perform contact-less melting experiments at high temperatures (>3000 K) and synthesize homogeneous glasses (Badro et al. 2021). Samples are placed on top of a conical nozzle and heated using an infrared $CO_2$ laser, and a gas stream flowing vertically from underneath the sample causes the sample to levitate, creating a controlled 1-bar atmosphere around the sample. A major advantage of this set-up is that the sample does not come into contact with any sample holder, which avoids contamination or unwanted chemical reactions between the sample holder and the sample. For example, with the gas-mixing furnace, while platinum wires are commonly used to suspend the samples, if a sample contains iron, a different wire material like rhenium must be used to minimize loss of iron from the melted sample to the metal wire. The ALLF can synthesize compositionally diverse melts, from primitive terrestrial magma analogs and metal-rich meteoritic analogs to synthetic exoplanet compositions thanks to its extremely fast quench rates (∼1000-1300 K/s) achieved by turning off the $CO_2$ laser. In addition, various gases and gas mixtures can be used to levitate the sample (e.g., Ar, $CO_2$, $H_2$, $O_2$, $N_2$, $SO_2$, $Cl_2$), establishing a range of atmospheric redox states surrounding the sample. ALLF systems are utilized to not only synthesize diverse melts but also to perform volatile solubility experiments for a broad range of melt compositions, temperatures and redox states under 1-bar total pressure conditions (e.g., Sossi et al. 2023).

At ETH Zürich, an ALLF system is coupled to an FTIR spectrometer covering wavelengths from ∼0.6 to 30 μm, providing novel, in-situ measurements of the outgassing composition from high-temperature melt samples (Fig. 6). The FTIR has two windows: one that uses an IR source with an external detector to measure the transmission spectra of gases evolving from the sample (Fig. 6 a, b) and one for measuring the emission spectra of outgassed species (Fig. 6 c). Importantly, this system covers the same wavelengths as JWST and will provide fundamental experimental data of the outgassing compositions and high-temperature gas opacities from compositionally diverse exoplanet melt analog materials over a range of temperatures and atmospheric redox states.

In addition to measuring volatile solubilities and outgassing compositions in diverse melt compositions at high temperatures, it is also important to place thermodynamic constraints on the activities of various melt species, which, under equilibrium conditions, are proportional to the partial pressures of these species in the vapor phase. Knudsen effusion mass spectrometry (KEMS) is a commonly used experimental technique for simulating equilibrium evaporation conditions and directly measuring vapor pressures of various molecules and thereby their activity coefficients. A KEMS system consists of a Knudsen cell which is an enclosed cell containing a condensed phase and vapor in equilibrium with a small orifice at the top of the cell that directs a molecular beam (i.e., directed flow) of the vapor phase to a mass spectrometer (Jacobson et al. 2024). While this technique was commonly used in the 1960-1970s, a resurgence is needed for exoplanet science to provide critical thermodynamic data on evaporation and solubility of volatiles in diverse melts representative of exoplanet analog materials. A new KEMS system is currently being built at the Carnegie Institution for Science's Earth and Planets Laboratory, led by PI A. Shahar, and will have the ability to heat materials up to ∼2700 K and measure the vapor pressures (and isotopic compositions) of evaporated species under equilibrium conditions for a wide array of planetary materials (Shahar 2024). This system will provide critical data on the activity coefficients for mantle materials and the evaporation composition from diverse planetary melts, which is essential for properly interpreting the outgassing atmospheres of low-mass exoplanets.

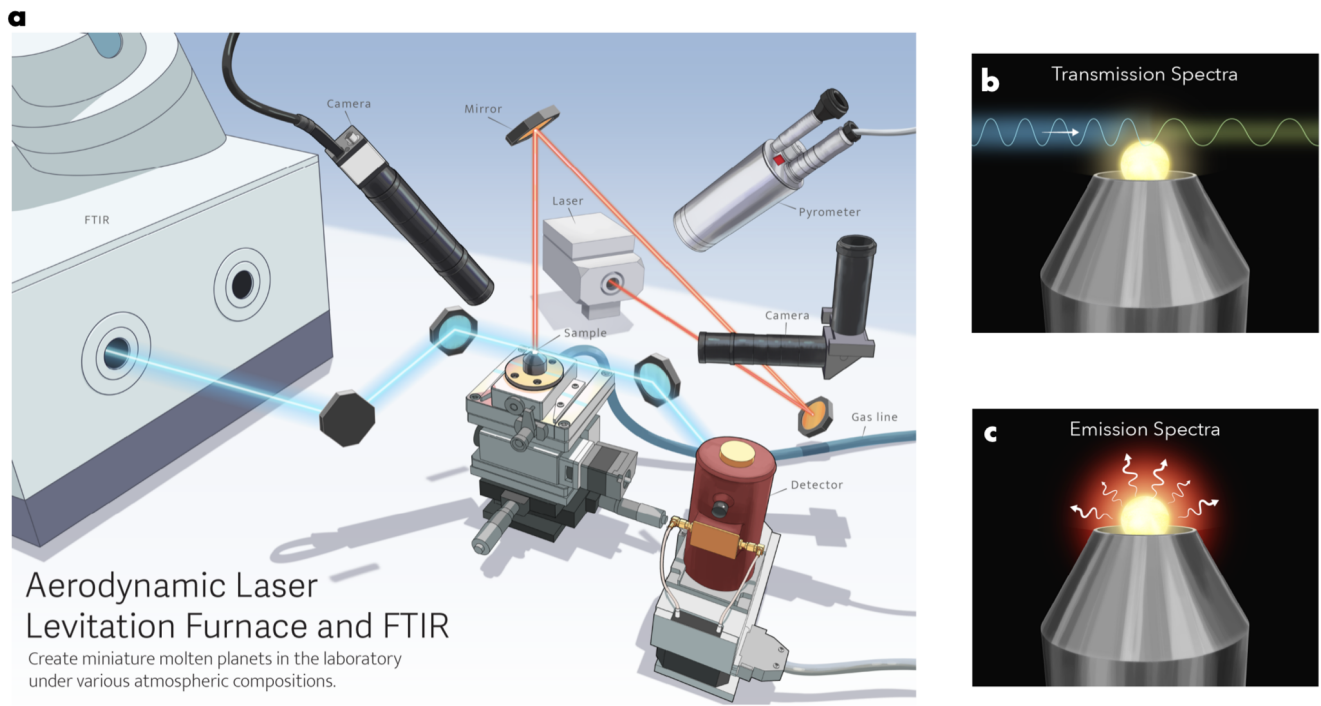

**Fig. 6** Schematic of the aerodynamic laser levitation furnace (ALLF) and FTIR spectrometer system at ETH Zürich's Experimental Planetology Group led by P. Sossi (a) and illustrations showing the two FTIR measurement modes: transmission (b) and emission (c). A solid sample is placed in the conical nozzle and levitated via a gas flow that is directed upwards from underneath the sample. The sample is heated using a $CO_2$ laser which is guided by a series of mirrors to heat the sample from the top (as indicated by the orange lines). Two cameras image the sample, and a pyrometer monitors the sample's temperature during the experiment. The FTIR spectrometer is used to measure either transmission (b) or emission (c) spectra of the species evaporating from the sample during the experiment. The blue line in (a) shows the FTIR measuring the transmission spectrum of the sample during an experiment. (© 2025 Elena Hartley http://www.elabarts.com)

### 2.5 High pressure-temperature experiments to constrain exoplanet interiors

Transit and radial velocity observations provide constraints on an exoplanet's radius and mass, respectively, from which the planet's bulk density can be derived. Exoplanet bulk densities have been used to infer interior compositions, but these two properties are inherently degenerate since multiple interior structures and mineralogies can result in the same bulk density (e.g., Valencia et al. 2007; Rogers and Seager 2010). To reduce these degeneracies, prior works have incorporated stellar compositions (e.g., Dorn et al. 2015, 2017) but directly linking an exoplanet's mass and radius to its bulk composition still requires further constraints from atmospheric observations. Any planet's interior composition and structure is governed by the equations of state (EOS) of its mineralogical phases which describe how the volume (or density) of a material varies with pressure and temperature (Baumeister et al. 2025). Many EOS are based on laboratory experiments that measure the volumes or densities of materials at high pressures and/or temperatures. There are a variety of experimental techniques used to simulate the high P-T conditions of planetary interiors, including static compression techniques such as piston-cylinder and multi-anvil presses and diamond-anvil cells (Fig. 1) along with dynamic compression methods using light gas guns, pulsed power and laser ablation techniques (for a review of these experimental methods see Fei and Tracy 2025). These methods are also essential for understanding how various elements partition between different phases. For example, for rocky planets, quantifying the partition coefficients for various elements between the core and mantle are critical for correctly modeling planetary interior structure and composition.

These high P-T experimental techniques are needed to determine volatile solubilities at the expected conditions of the magma ocean-envelope boundary for sub-Neptune ex-oplanets. For example, recent works by Miozzi et al. 2025 and Horn et al. 2025 used laser-heated diamond-anvil cell experiments to constrain water solubility at pressures and temperatures up to 60 GPa and 4000 K, respectively (Horn et al. 2025; Miozzi et al. 2025). Both studies find that large amounts of hydrogen can dissolve in silicate melts at these high P-T conditions and produce water which will influence the interior structures of sub-Neptunes exoplanets. To properly interpret current and upcoming observations of sub-Neptunes like K2-18 b and TOI-270 d, we need further experimental constraints on both the equations of state of various mineral phases along with solubilities of atmospheric gases at these high P-T conditions.

Despite great advances in these experimental methods in recent years (e.g., Duffy and Smith 2019), reaching the extremely large pressures and temperatures expected in the deep interiors of some sub-Neptune exoplanets is beyond the feasibility of current laboratory techniques. Therefore, computational techniques must be utilized such as ab initio calculations that determine the high P-T EOS and thermal properties of materials from first-principles quantum mechanics (e.g., Caracas 2024; Young et al. 2025). A commonly used ab initio method is density functional theory molecular dynamics (DFT-MD) simulations which compute the structure and thermodynamic properties of materials at the atomic level. As possible, it is important to compare the outputs of ab initio simulations to experimental data to help ground-truth the computational calculations.

## 3 Spectroscopic data and atmospheric chemistry

The ability to properly interpret exoplanet atmospheric observations relies heavily on the quality of our atmospheric models which depends both on the treatment of various atmospheric processes and the associated input data. The physical and chemical data necessary for modeling planetary atmospheres is primarily associated with how vapor and condensed species interact with radiation and each other, including properties like gas opacities, thermo- and photochemical reactions and their rates, and aerosol compositions and structures. This data derives in large part from laboratory experiments, which, for exoplanet science, must expand to conditions unlike those in Earth's modern atmosphere (e.g., higher temperatures and pressures, different compositions and radiation environments). This section focuses on laboratory measurements used to obtain spectroscopic data and radiative transfer properties (e.g., opacities, cross sections) and constrain atmospheric chemistry (both thermochemistry and photochemistry) for gaseous species and mixtures. Section 4 will discuss the experimental methods associated with simulating and characterizing atmospheric aerosols, both condensate clouds and photochemical hazes.

### 3.1 Spectroscopic data of atmospheric gases

Opacity is a measure of a medium's impenetrability to radiation. Atmospheric opacity is wavelength-dependent and depends on the gas molecules present, their abundances and absorption cross sections (i.e., the probability that a molecule will absorb radiation), the atmospheric pressure-temperature conditions and density, and any scattering and collision-induced processes. A comprehensive understanding of opacities is necessary for modeling radiative transfer and simulating the spectrum of a planet's atmosphere (Chance and Martin 2017). Since atmospheres are composed of mixtures of various gas species, interactions between gas molecules can impact the shape of spectral lines and thereby their opacities. For example, the dominant gases in an atmosphere can collide with radiatively active molecules which induces broadening of their spectral lines (i.e., pressure/collision broadening) (Rayer 2020). While pressure broadening parameters are well-understood for Earth's atmosphere, there is a need for more experimental data on these parameters for atmospheres dominated by $H_2$ and He, $N_2$ and/or $O_2$, $CO_2$, and $H_2O$ at temperature and pressure conditions unlike those in modern Earth's atmosphere (i.e., >1 bar and temperatures spanning ∼500-4000 K) (Fortney et al. 2016, 2019).

In dense regions of a planetary atmosphere, collisions can also induce continuum absorption, both from induced transitory dipole moments in gas molecules (i.e., collision-induced absorption (CIA)) and from the build-up of supramolecules (e.g., dimers, trimers, etc.) that create broad absorption features (Richard et al. 2012; Karman et al. 2019). Continuum absorption parameters have been well-studied for Earth's atmosphere and other Solar System bodies including gas giants (e.g., CIA for $H_2$, He and H), Titan (e.g., $N_2$, $CH_4$, $H_2$ CIA), and Venus (e.g., $CO_2$-$CO_2$ continuum absorption). However, there are still gaps in our understanding of CIA and dimer (or larger supramolecular) absorption for different sets of molecules and under conditions relevant for exoplanet atmospheres (Fortney et al. 2016). More laboratory data on continuum opacities is required to fill these gaps. In addition, exoplanet atmosphere observations are utilizing high-resolution spectroscopy which requires that molecular opacity data be properly calibrated at such resolutions and the temperature and pressure conditions relevant for these planets' atmospheres to ensure correct analysis of the observations.

The experimental instruments for measuring absorption cross sections and spectral line shapes and transitions often consist of a radiation source coupled to a cell or cavity containing the gas or gas mixture which is also connected to a spectrometer such as a Fourier Transform Spectrometer (FTS) or echelle spectrometer. It is important to note that sometimes the radiation source is contained within the spectrometer, such as a globar source in an FTIR. High-resolution FTIR instruments can resolve individual lines and determine pressure/collision-induced broadening coefficients (e.g., Turbet et al. 2019, 2020; Hendaoui et al. 2025). For example, recent works by Turbet et al. used a high-resolution FTIR connected to a multi-pass cell to measure far-IR (∼15-250 μm) absorption by $CO_2$, $CH_4$, $H_2$ and $CO_2+CH_4$ and $CO_2+H_2$ mixtures at room temperature (Turbet et al. 2019). Using these laboratory measurements, Turbet et al. derived a semi-empirical model to compute $CO_2+CH_4$ and $CO_2+H_2$ CIAs across broader spectral and temperature ranges and used this model to better understand Mars' early atmosphere and its greenhouse warming (Turbet et al. 2020). Recent work by Ranjan et al. used a gas cell coupled with a far-UV spectrometer to measure near-UV (>200 nm) $H_2O$ vapor absorption cross sections at 292 K. They found that $H_2O$ absorbs much more than previously assumed, and these higher cross sections cause enhanced OH production which influences photochemical modeling and biosignature interpretation for low-mass planet atmospheres (Ranjan et al. 2020). Another commonly used method for measuring absorption line transitions and intensities is cavity ring-down spectroscopy (CRDS) which consists of a pulsed laser beam that enters an optical cavity with two highly reflective mirrors and bounces back and forth. The intensity of light that is transmitted through the mirror is measured by a detector over time which provides the rate of absorption by the medium in the optical cavity and thereby allows direct determination of line intensities, shape parameters and absorption coefficients (Leshchishina et al. 2012; Gupta 2020; Adkins et al. 2021; Maity et al. 2021).

There are various databases that compile spectroscopic parameters (e.g., line-by-line parameters, absorption cross sections, CIAs) for atmospheric modeling including HITRAN (and its high-temperature extension HITEMP) (Rothman 2021), GEISA (Jacquinet-Husson et al. 2008), PNNL Spectral Library (Johnson et al. 2004), ExoMol (Tennyson et al. 2016), the JPL Molecular Spectroscopy Catalog (Pickett et al. 1998), and CDMS (Müller et al. 2001). These databases incorporate experimental data either directly or to validate theoretical calculations. While ab initio calculations are used to predict various spectroscopic properties (e.g. line shape parameters, CIAs) of gas molecules and mixtures at extreme conditions relevant for exoplanet atmospheres, they importantly rely on laboratory data at ambient conditions to validate these calculations. Future laboratory efforts should focus on measuring absorption properties for various gases (and gas mixtures) at temperatures relevant for exoplanet atmospheres (up to ∼3000 K) (Niraula et al. 2022).

## 3.2 Chemical reactions and rates

In addition to spectroscopic data, a comprehensive understanding of planetary atmospheres requires knowledge of the chemical reactions that take place. Two primary types of chemical reactions occur in planetary atmospheres: 1) thermochemical reactions driven by temperature and collisions between gas molecules and 2) photochemical reactions initiated by gas molecules absorbing photons (Chamberlain and Hunten 1987; Moses 2014; Madhusudhan et al. 2016). Both types of chemical reactions influence the structure and composition of planetary atmospheres, with photochemistry mainly affecting the upper levels where UV photons are abundant (Yung and DeMore 1999) and thermochemistry dominating in deeper, hotter regions (Hu and Seager 2014). Laboratory data is essential to constrain thermo- and photochemical reaction rates at the pressure-temperature conditions in planetary atmospheres. Experimental set-ups used for measuring chemical reaction rates are similar to those used to obtain spectroscopic data but can involve other types of spectrometers (e.g., mass spectrometry, laser magnetic resonance spectrometry) and involve various radiation sources (e.g., VUV lamps) for generating photochemical reactions.

The high-temperature and/or high-pressure chemical reaction data that is needed to understand gas giants, magma planets and rocky worlds with thick atmospheres such as Venus, derives in large part from the combustion literature and requires laboratory measurements from combustion reactors and shock tube laboratories (e.g., (Bauer 1963; Tranter et al. 2001; Davidson and Hanson 2009; Hanson and Davidson 2014; Winiberg et al. 2025). For example, a recent study by Panda et al. utilized a shock tube with laser absorption spectroscopy at Stanford's Kinetic Shock Tube Facility to determine sulfur chemical reaction rates, in particular the rate coefficient of OCS decomposition, under various background gases at several pressures (1-8 bars) and high temperatures (1800-2500 K) (Panda et al. 2026). Such experimental constraints on chemical reaction rates for sulfur species are important for properly modeling sulfur kinetics and its impact on atmospheric chemistry for Venus-like planets, as highlighted by Wogan et al. 2025.

Chemical reaction rate coefficients are compiled in multiple databases such as the Kinetic Database for Astrochemistry (KIDA, Wakelam et al. 2012), the NIST Chemical Kinetics Database (Mallard et al. 1998), the JPL Kinetics database (Burkholder et al. 2019), and the International Union of Pure and Applied Chemistry (IUPAC) Task Group on Atmospheric Chemical Kinetic Data Evaluation (Atkinson et al. 2004). However, more laboratory experiments along with new laboratory techniques are needed to determine chemical reaction rates at the diverse temperature, pressure and compositional conditions expected in exoplanet atmospheres (Fortney et al. 2016; Stapelfeldt and Mamajek 2025; Winiberg et al. 2025).

## 4 Exoplanet aerosols

Small, suspended solid or liquid particles are a common component of planetary atmospheres within and beyond the Solar System. As in Gao et al. 2021, we will use the term "aerosols" to encapsulate all kinds of particles found in planetary atmospheres. For exoplanet atmospheres, clouds and hazes are the two primary types of aerosols, with clouds consisting of collections of particles that form under thermochemical equilibrium (e.g., phase changes and other chemical reactions) and hazes, which originate directly from energy input via photochemistry and particle bombardment (i.e., locally irreversible processes) (Gao et al. 2021). Aerosols have a strong impact on the observability of exoplanet atmospheres because their opacities block atmospheric regions below cloud or haze layers. In addition, aerosols influence the thermal structure and albedo of a planet's atmosphere, and their scattering and absorption properties generate atmospheric features.

While a detailed understanding of the composition and structural properties of exoplanet clouds is lacking, combined observational and theoretical works on the role of clouds in brown dwarf and hot Jupiter atmospheres have made significant progress over the last several decades. For example, the formation and evolution of refractory clouds composed of Ti, Va, Fe and silicates characterize the transitions between brown dwarf spectral types (e.g., Burrows and Sharp 1999; Lodders and Fegley 2002; Marley et al. 2002; Kirkpatrick 2005). In addition, recent studies of hot Jupiter exoplanets find that the presence of silicate clouds are consistent with spectroscopic observations and can explain the thermal properties of their atmospheres (e.g., Gao and Powell 2021; Inglis et al. 2024). Laboratory measurements of the optical properties of potential exoplanet cloud condensates use similar experimental set-ups as those discussed in Sect. 3 for obtaining spectroscopic data of atmospheric gases. Various works have compiled optical constants for sets of cloud condensates, which derive in large part from experimental measurements (e.g., Wakeford and Sing 2015; Kitzmann and Heng 2018). Future efforts should aim to experimentally constrain the optical and structural properties of exoplanet cloud analogs at the temperature, pressure and irradiation conditions anticipated for exoplanets.

As evidenced by Earth's atmosphere over time and that of Saturn's moon Titan, photochemical hazes can also have a strong influence on a planet's atmospheric composition and structure. Titan's atmosphere currently has an organic-rich haze layer, and significant laboratory efforts have been devoted to understanding the formation and composition of these hazes. These laboratory-based Titan haze analogs are often referred to as "tholins" (Cable et al. 2012; Hörst 2017). During Earth's Archean eon (∼4-2.5 Ga), it is possible that a similar organic-rich haze layer formed due to the higher levels of atmospheric $CH_4$ which largely derived from biological activity and would have influenced early Earth's surface temperature (e.g., Arney et al. 2016; Catling and Zahnle 2020). As a result, the presence of an organic haze has been proposed as a possible biosignature on terrestrial exoplanets (Arney et al. 2018). Significant theoretical work has been done to model atmospheric clouds and hazes both for Solar System bodies and exoplanets, ranging in complexity from simple parameterizations to microphysical models ranging from 1D to 3D (e.g., Ackerman and Marley 2001; Greene et al. 2016; Line and Parmentier 2016; Gao and Benneke 2018; Powell et al. 2019). Laboratory measurements are essential for constraining these models and properly determining the effects of photochemical hazes in exoplanet atmospheres.

The experimental methods used to generate photochemical haze analogs consist of a temperature-controlled reaction chamber containing a gas mixture that is coupled to a radiation source, which is typically a UV lamp or plasma discharge (Fig. 1). Some set-ups couple the reaction chamber with a detector such as a mass spectrometer to monitor the gas composition during the experiment. The reaction chamber usually contains optical windows or other surfaces that act as substrates for collection of the haze material. After the experiment, the haze is then analyzed using various instruments including mass spectrometers to determine the chemical composition, other spectrometers to measure optical properties from the UV to the IR (e.g., FTIR spectrometers, spectrophotometers), pycnometer to measure density, and atomic force microscopy (AFM) to determine particle sizes. Examples of haze generating laboratories around the world include the Planetary Haze Research (PHAZER) chamber at Johns Hopkins University, USA (Hörst et al. 2018; He et al. 2019), the Production d'Aérosols en Microgravité par Plasma REactifs (PAMPRE) at the Laboratory for Atmospheres, Observations and Space (LATMOS), France (Szopa et al. 2006), the Cosmic Simulation Chamber (COSmIC) at NASA Ames Research Center, USA (Sciamma-O'Brien et al. 2014) (Sciamma-O'Brien et al. 2014) and the Photochemical Aerosol Chamber (PAC) at the University of Northern Iowa, USA (Sebree et al. 2018).

Recent experimental works are simulating exoplanet hazes by exploring regions of the gas composition, temperature, and irradiation environment parameter space outside of those relevant for Titan or Earth and measuring various properties including their chemical compositions, structures and optical properties. For example, a series of papers by He et al. simulated photochemical haze formation using PHAZER for a range of atmospheric metallicities and temperatures. They determined how the particle size distributions of the hazes vary with temperature and metallicity and measured the hazes' optical properties and densities (He et al. 2019, 2023). A complementary study by Moran et al.

measured the chemical compositions of PHAZER-generated exoplanet haze analogs using high-resolution mass spectrometry, finding that complex organic molecules form including some with prebiotic potential (Moran et al. 2020). In addition, Yu et al. measured the surface energies of these hazes, providing important constraints on their removal efficiencies from the atmosphere (Yu et al. 2021). Drant et al. 2025 performed a cross-laboratory study that measured the refractive indices from the UV to the near-IR of photochemical haze analogs produced by two different laboratories, PAMPRE and COSmIC (Drant et al. 2025). This study found that the specifics of the experimental set-up, such as the gas residence time, irradiation method, pressure and temperature conditions and the $N_2/CH_4$ ratio of the gas influence the composition and optical properties of the haze. Additional experimental work is needed to constrain the diverse compositions, structures and optical properties of photochemical hazes and how they influence exoplanet atmosphere observations. We also need more cross-laboratory studies like Li et al. 2022 and Drant et al. 2025 to ensure that the differences between experimental set-ups are well-understood and the effects of those differences are quantified (Li et al. 2022; Drant et al. 2025).

## 5 Exoplanet surfaces

At present, we do not have a definitive detection of an atmosphere around a rocky exoplanet. While JWST has observed dozens of rocky worlds and super-Earths via transmission spectroscopy such as TRAPPIST-1 b and c, L 98-59 c, TOI-836 b, and GJ 341 b, the data are all consistent with flat lines (Greene et al. 2023; Zieba et al. 2023; Alderson et al. 2024; Kirk et al. 2024; Scarsdale et al. 2024). These observations only allow us to rule out hydrogen-dominated atmospheres but cannot be distinguished from a cloudy atmosphere, one with high mean molecular weight, or an air-less body (Kreidberg and Stevenson 2025). Perhaps the best evidence for an atmosphere around a rocky exoplanet is from recent JWST observations of the emission spectrum of the ultra-short period (P < 1 day) magma planet TOI-561 b (Teske et al. 2025). These observations find that the planet's dayside is inconsistent with a bare-rock surface, suggesting that it has a thick volatile atmosphere to cool the surface below what would be expected if it had no or a very thin atmosphere. In the coming years, more JWST observations pushing to higher precision are expected to uncover the presence of atmospheres on rocky and super-Earth exoplanets. Nevertheless, it is likely that some portion of rocky worlds do not have atmospheres or have very tenuous ones. For example, Spitzer observations of the phase curve of rocky exoplanet LHS 3844 b are consistent with a bare-rock that has a basaltic surface composition (Kreidberg et al. 2019).

Constraining the chemical compositions of exoplanet surfaces requires a comprehensive understanding of the optical properties of diverse surface materials. Laboratory measurements are essential for providing reference spectroscopic data of exoplanet surface analogs that are necessary for interpreting observations. There are multiple reflectance and emission spectral libraries of rocky materials such as the Reflectance Experiment Laboratory (RELAB) database at Brown University (Milliken et al. 2021) and the ASU Thermal Emission Spectroscopy Laboratory Spectral Library (Christensen et al. 2000). A recent study by Hammond et al. utilized the spectral properties for a diverse set of rocky surface materials from the RELAB database to simulate JWST MIRI observations and determine if bare-rock planets can be distinguished from those with atmospheres. They find that secondary eclipse observations can be degenerate between surfaces and atmospheres and suggest that instead phase curve observations are needed to distinguish between bare-rock planets and those with atmospheres (Hammond et al. 2025). Recent experimental studies have synthesized possible exoplanet surface materials and measured their spectral and photometric properties (e.g., Fortin et al. 2022; First et al. 2025; Paragas et al. 2025). For example, Fortin et al. created over a dozen exoplanet surface analog materials based on plausible exoplanet mantle compositions derived from host star chemical abundances from Putirka and Rarick 2019 along with several Solar System-based materials and measured their infrared reflection spectra from 2.5-28 μm using FTIR and Raman spectroscopy (Putirka and Rarick 2019; Fortin et al. 2022). Based on their measurements, Fortin et al. established a relationship between the surface composition and the location of the Christiansen spectral feature near ∼8 μm. Another study by First et al. measured the emissivities of basaltic samples from 2-25 μm using FTIR and then used their experimental data to simulate JWST spectra of bare-rock exoplanets with these basaltic surfaces to determine if JWST can distinguish between different surface compositions (First et al. 2025). They conclude that JWST's mid-IR instrument (MIRI) should be able to distinguish and observe several mineralogical and chemical signatures, especially hydrous minerals (e.g., serpentine and amphibole). Many of these measurements involved using an integrating sphere with the spectrometer since these materials can have a high degree of scattering.

It is also important to determine the spectroscopic properties of these materials over a range of temperatures. Biren et al. 2022 measured the emissivities of basalts in-situ from room temperature to 1800 K using FTIR spectrometers coupled to a sample chamber with a $CO_2$ laser heating source and found that emissivity from ∼1-29 μm varies with temperature and the material's degree of polymerization (Biren et al. 2022). A recent study by Paragas et al. measured the emissivities of surface analog materials with various textures (e.g., solid slab, fine powder) over a range of temperatures from 500-800 K using an emissivity chamber coupled to an FTIR (Paragas et al. 2025). They then incorporated these experimental results into exoplanet retrievals. While they find that the surface material's texture can have a significant impact on a planet's albedo, they note that temperature-dependent changes in surface spectral features will likely be undetectable at the current precision of observations. Additional experimental work is needed to constrain the spectroscopic properties of exoplanet surface analog materials across a range of bulk compositions, textures and temperatures.

# 6 Astrobiology experiments for biosignature interpretation

One of the main endeavors for exoplanet science over the coming decades is to search for signs of life on rocky, potentially habitable planets. A common first-order means of identifying plausible habitable exoplanets is to search for ones in the Habitable Zone (HZ) of their host star, which is defined by the range of distances around a star that permit a planet with an atmosphere to maintain surface liquid water (Kasting et al. 1993; Kopparapu et al. 2013). However, a complete assessment of planetary habitability requires understanding the complex interplay among many factors, both stellar and planetary, including instellation from the host star, orbital architecture and dynamics of the planetary system, atmospheric composition and properties, planetary tectonic regime, surface-atmosphere and surface-interior feedback effects (Kopparapu et al. 2020; Krissansen-Totton et al. 2022).

One of the primary challenges to searching for life on rocky exoplanets is that our present knowledge of the possible diversity of biochemistries that may exist is severely limited to a sample size of one. Therefore, our current understanding of life's requirements is confined only to our knowledge of life on Earth, both at present and throughout Earth's history. Based on this, life requires an energy source (i.e., light and redox chemical energy), a supply of elements that can form molecules with a diversity of shapes and properties (i.e., carbon, hydrogen, nitrogen, oxygen, phosphorus, sulfur, collectively referred to as CHNOPS), a solvent to support the synthesis and interaction of these molecules (i.e., liquid water), and the physical and chemical conditions to support the molecules and interactions that biochemistry needs (Hoehler et al. 2020). As we search for signs of life on exoplanets, it is essential to generalize our understanding of life's requirements and the various ways in which life can modify a planet's environment.

Chemical and/or physical indications of present or past forms of life are referred to as "biosignatures," and, to date, much of the focus has been on determining the planetary context for various atmospheric biosignatures that could be detected with current and upcoming space-based telescopes like JWST. Commonly explored biosignature gases include $O_2$ and its photochemical byproduct $O_3$ (e.g., Meadows et al. 2018), $CH_4$ (e.g., Thompson et al. 2022), $N_2O$, $NH_3$, and various hydrocarbons like $CH_3Cl$ and $C_2H_6$ (e.g., Schwieterman et al. 2018). When evaluating possible biosignature gases, it is important to not only consider how present biological activity generates these gases but also how they are produced abiotically to rule out false positive scenarios and to account for how they were generated, both biotically and abiotically, throughout Earth's history. In addition, the coexistence of multiple gases can serve as biosignature pairs or groups which may modify a planet's atmosphere in observable ways (e.g., (Krissansen-Totton et al. 2018). For example, organic haze generated by biogenic $CH_4$ and organic sulfur gases could serve as a biosignature for anoxic (i.e., oxygen-depleted) Earth-like atmospheres (Arney et al. 2018).

Beyond gaseous biosignatures, surface biosignatures have been proposed and include both photosynthetic and non-photosynthetic pigments. For example, photosynthetic pigments due to the cellular structure of vegetation can be detected spectroscopically, including the strong reflectance of Earth plants in the near-IR (∼700-750 nm) called the "vegetation red edge" (Seager et al. 2005). While non-photosynthetic pigments do not capture light, they nevertheless interact with it incidentally and can also produce observable spectral features (Schwieterman et al. 2015). Another form of surface biosignatures include polarization biosignatures. For example, a key property of all life on Earth is homochirality (i.e., single-handedness), and when biological compounds interact with unpolarized light, homochirality induces fractional circular polarization in the light that gets scattered from a biological compound. Various studies have assessed how chirality can be observed as a surface biosignature via spectral detections of circular polarization signatures using laboratory spectral measurements of surface microbes and vegetation species (e.g., Sparks et al. 2021; Patty et al. 2022). Photosynthetic pigments can also linearly polarize light, and laboratory efforts such as the bio-polarimetric laboratory BioPol determine the optical polarized spectra of various biotic and abiotic samples, which is critical for assessing potential surface biosignatures (Berdyugina et al. 2016).

Laboratory experiments provide an important mechanism to explore ways in which life can modify exoplanet surfaces and atmospheres in spectroscopically observable ways and to determine how potential biosignatures behave in environments unlike those of present-day Earth. To comprehensively understand how possible biosignature gases like $O_2$ and $CH_4$ can accumulate in exoplanet atmospheres, the atmospheric chemistry experimental set-ups discussed in Sects. 3 and 4 can be used to simulate different irradiation environments and atmospheric compositions. For example, the Berlin Atmospheric Simulation Experimental chamber (BASE) is a new laboratory facility for directly studying atmospheric chemistry relevant for exoplanets, including in-situ monitoring of the abundances of proposed biosignature gases, under diverse irradiation environments, which will be useful for identifying false positive scenarios (Hofmann et al. 2024). In addition to experimental constraints on how proposed biosignature gases may accumulate abiotically in planetary atmospheres to generate false positive signals, it is imperative to assess both the plausible source and sink mechanisms for such gases. Such an assessment requires modeling efforts combined with laboratory experiments.

While the vast majority of life on Earth today is produced via oxygenic photosynthesis, during the Archean eon (4-2.5 Ga) prior to the rise of oxygen in Earth's atmosphere and oceans, the atmosphere was anoxic, and life utilized anoxygenic metabolisms such as methanogenesis (Wolfe and Fournier 2018; Thompson et al. 2022). If life is common in the universe, it is possible that methanogenesis and other simpler anoxygenic metabolisms may be widespread due to the likely ubiquity of the $CO_2$-$H_2$ redox couple in terrestrial planet atmospheres and the ability to exploit commonly outgassed substrates (Thompson et al. 2022). A laboratory study by Coehlo et al. measured the reflectance spectra of anoxygenic phototrophs (i.e., purple bacteria) and used their data to model the spectra of rocky exoplanets with surfaces dominated by purple bacteria (Coelho et al. 2024).

Recent experimental efforts are examining how life can be observed in planetary atmospheres. For example, a study by Coehlo et al. measured the reflectance spectra for a diverse set of biopigments produced by microbes found in Earth's atmosphere, and they propose that such biopigments produced by aerial bacteria can serve as biosignatures in exoplanet atmospheres (Coelho et al. 2025). Other studies are also considering how life can exist in non-Earth-like environments. For example, it is likely that some portion of low-mass exoplanet atmospheres are dominated by $H_2$, and Seager et al. experimentally studied how single-celled microorganisms respond to being in a pure $H_2$ atmosphere. By monitoring cell cultures exposed to $H_2$ gas over time, they found that both E. coli and yeast can survive and grow in liquid cultures in a pure $H_2$ atmosphere and that E. coli produces various gases in such an environment, including several proposed biosignatures like $N_2O$, $NH_3$, and dimethyl-sulfide (DMS) (Seager et al. 2020). While liquid water is the primary solvent used by life on Earth, experimental studies are also considering if alternative solvents may be utilized by life on other planets. For example, Agrawal et al. experimentally demonstrated that for planets with thin atmospheres, ionic liquids can form on surfaces from sulfuric acid and nitrogen-containing organic compounds and may serve as a solvent (Agrawal et al. 2025).

In preparation for next-generation missions like the Habitable Worlds Observatory and the Large Interferometer for Exoplanets (LIFE), additional astrobiology experiments that determine how proposed biosignature gases can be produced biotically or abiotically in different atmospheric and irradiation environments will be essential to ensure that upcoming observations are properly interpreted (Quanz et al. 2022; Harada et al. 2024). At the same time, experiments that determine how life can exist and grow under diverse surface and atmospheric conditions and with alternative solvents are similarly important for these upcoming missions.

## 7 Conclusions

This review highlights the diversity of laboratory techniques that are used to obtain fundamental data to constrain key properties of exoplanet atmospheres, surfaces and interiors, which is necessary for properly interpreting spectroscopic observations of these worlds (Fig. 1). For the foreseeable future, remote observations of exoplanet atmospheres (and surfaces for airless planets) will be the main window for studying planets beyond our Solar System and for determining their chemical and physical properties (Crossfield 2015). To properly interpret observations of an exoplanet's atmosphere and surface, laboratory data is needed to provide inputs on the spectroscopic properties of atmospheric gases and aerosols along with surface analog materials (Fortney et al. 2019). One of the key goals of exoplanet science is to link spectroscopic observations of an exoplanet's atmosphere (or surface) to properties of its interior. In this review, we discuss recent experimental works that obtained data on how essential atmosphere-forming volatile elements (e.g., H, O, C, S) partition between atmospheres and interiors (e.g., outgassing and solubility experiments) over a broad range of pressure, temperature and redox conditions, which are important constraints for coupled interior-atmosphere models that are used to connect atmospheric observations to interior properties (e.g., Thompson et al. 2021, 2023, 2025; Bower et al. 2025; Miozzi et al. 2025).

Looking ahead to next-generation space missions like the Habitable Worlds Observatory and LIFE, the driving science goal will be to search for signs of life on habitable zone, rocky exoplanets (Quanz et al. 2022; Harada et al. 2024). To achieve this goal, laboratory data on the behavior of possible biosignature gases in environments unlike that of modern Earth and how life can modify a planet's atmosphere or surface in spectroscopically observable ways will be essential to assess a potential sign of life on an exoplanet. Definitively detecting a sign of life on an exoplanet will be one of the most revolutionary scientific discoveries in human history. Therefore, it is important that the exoplanet community prioritizes establishing ground-truth laboratory datasets to ensure that such observations are properly interpreted.

**Acknowledgements** M.A.T thanks the anonymous reviewer whose comments helped to strengthen the article. M.A.T acknowledges Myriam Telus, Jonathan Fortney, Joshua Krissansen-Totton, and Laura Schaefer for their guidance while serving on the PhD dissertation committee for the thesis which presented some of the results discussed in this manuscript. M.A.T. also acknowledges Paolo Sossi and Dan J. Bower for their guidance on several of the recent papers reviewed in this manuscript.

**Author contributions** M.A.T performed the research, wrote the main papers reviewed in this manuscript (i.e., Thompson et al. 2021, 2023, 2025) and wrote this review manuscript.

**Funding information** This work was supported by NASA through the NASA Hubble Fellowship grant #HST-HF2-51545 awarded by the Space Telescope Science Institute, United States, which is operated by the Association of Universities for Research in Astronomy, Inc., for NASA under contract NAS5-26555.

**Data availability** No datasets were generated or analysed during the current study.

## Declarations

**Ethics declaration** Not applicable.

**Competing interests** The authors declare no competing interests.